# Interpretation of some Yb-based valence-fluctuating crystals as approximants to a dodecagonal quasicrystal

Tsutomu Ishimasa [a*], Marek Mihalkovič [b], Kazuhiko Deguchi [c], Noriaki K. Sato [c] and Marc de Boissieu [d]

[a] *Toyota Physical and Chemical Research Institute, Nagakute, Aichi 480-1192, Japan;*
[b] *Institute of Physics, Slovak Academy of Sciences, 845 11 Blatislava, Slovakia;*
[c] *Department of Physics, Graduate School of Science, Nagoya University, Nagoya 464-8602, Japan;*
[d] *University of Grenoble Alpes, CNRS, Grenoble INP, SIMaP, F-38000, Grenoble, France.*

**Abstract**

The hexagonal ZrNiAl-type (space group: $P\bar{6}2m$) and the tetragonal $Mo_2FeB_2$-type (space group: $P4/mbm$) structures, which are frequently formed in the same Yb-based alloys and exhibit physical properties related to valence-fluctuation, can be regarded as approximants of a hypothetical dodecagonal quasicrystal. Using Pd-Sn-Yb system as an example, a model of quasicrystal structure has been constructed, of which 5-dimensional crystal (space group: $P12/mmm$, $a_{DD}$=5.66 Å and $c$=3.72 Å) consists of four types of acceptance regions located at the following crystallographic sites; Yb [00000], Pd[1/3 0 1/3 0 1/2], Pd[1/3 1/3 1/3 1/3 0] and Sn[1/2 00 1/2 1/2]. In the 3-dimensional space, the quasicrystal is composed of three types of columns, of which $c$-projections correspond to a square, an equilateral triangle and a 3-fold hexagon. They are fragments of two known crystals, the hexagonal α-YbPdSn and the tetragonal $Yb_2Pd_2Sn$ structures. The model of the hypothetical quasicrystal may be applicable as a platform to treat in a unified manner the heavy fermion properties in the two types of Yb-based crystals.

# 1. Introduction

The concept of "approximant" introduced by Elser and Henley [1] established relationship of some complex but periodic structures to quasicrystals. The relationship stems from the identity of the building structural motifs/units that form periodic arrangements in the "approximant" crystals, and lack translational periodicity in the quasicrystal phase. In the framework of the projection method [2-4], the quasicrystal is described as an incommensurate projection of a higher-dimensional crystal onto "real" or "physical" subspace, while an approximant arises from slightly distorted hypercrystal; the corresponding distortion is called linear phason strain [5].

In the history of quasicrystal research, approximants served three important roles; firstly the existence of such crystal phase indicated possible existence of quasicrystal at a similar composition. Secondly it provided detailed and direct access to the structure of the building blocks shared between approximant and quasicrystal phases. And thirdly, whenever both quasicrystal and approximant phases were available, comparison of their properties highlighted features that were specifically due to the aspect of quasiperiodic long range order. The recent discovery of unconventional quantum criticality in the Au-Al-Yb quasicrystal is a typical example of the application of the relation between the quasicrystal and the corresponding approximant [6]. While the magnetic property of the approximant is sensitive to pressure and the quantum criticality emerges at the limiting pressure of around 2.0 GPa, the quantum criticality of the quasicrystal phase is insensitive to the pressure [7]. This difference possibly originates from the critical character of wave functions of electrons in the quasicrystal. Testing this hypothesis requires accumulating further knowledge in wider range of materials.

In this paper, we will treat two types of structures containing Yb. One is hexagonal ZrNiAl-type (space group: $P\bar{6}2m$) and the other tetragonal $Mo_2FeB_2$-type (space group: $P4/mbm$) structures. Such Yb-based crystals have been studied intensively for the last decade from the viewpoint of valence-fluctuation, and competition between the Kondo effect and the Ruderman-Kittel-Kasuya-Yosida interaction. The former, Kondo effect, favors a nonmagnetic ground state, and the latter magnetic ordered state. Competition of the two effects gives rise to intriguing phenomena such as heavy fermions, non-fermi liquid and quantum criticality. In the alloy system of Yb-Pd-Sn, both types of structures are formed [8, 9]. The hexagonal α-YbPdSn belonging to ZrNiAl-type is formed as a low-temperature phase around 600 °C [10]. This crystal exhibits valence-fluctuation, and also antiferromagnetic order at very low temperature, $T_N$=200 mK [11]. Recently, the isostructural YbAgGe was reported to show quantum bicriticality [12]. The tetragonal $Yb_2Pd_2Sn$ also shows interesting properties related to valence-fluctuation. This crystal exhibits pressure-driven, or composition-driven, two quantum critical points [13, 14]. In this structure Yb atoms are located at the vertices of triangle-square tiling, and the magnetic property of this type of structure was discussed from the viewpoint of geometrical frustration forming the Shastry-Sutherland lattice [15]. Although these two phases are frequently formed as adjacent phases in several alloy systems including rare-earth elements or actinides [16, 17], their physical properties have been discussed individually.

In this paper it will be shown that these two types of structures can be understood as approximants of a unique dodecagonal quasicrystal that has not been synthesized yet. The structure model may be useful in order to study physical properties of Yb-based valence-fluctuating crystals related to the quasicrystal.

## 2. Structural properties of α-YbPdSn and $Yb_2Pd_2Sn$

Figs. 1a and b present structures of α-YbPdSn and $Yb_2Pd_2Sn$, respectively. The former is hexagonal with the lattice parameters $a_h$=7.590 Å and $c_h$=3.770 Å [10]. The latter is tetragonal; $a_t$=7.589 Å and $c_t$=3.635 Å [8]. Both are layered structures with atoms located at $z$=0 and 1/2, and stacked along their respective $c$-axes with period agreeing to within 4%.

The $c$-projection of the hexagonal α-YbPdSn reveals that linkages between Yb atoms (in $z$=0 layer) separated by 4.01Å define periodic tiling of two tiles, equilateral triangle and the *shield* - a 3-fold hexagon. The triangles are centered by a Pd atom at $z$=1/2, while shields are decorated by Sn triangles in $z$=1/2 layer and centered by another Pd atom at $z$=0. There are two crystallographic sites for Pd that are named $Pd_1$ and $Pd_2$ as shown in Fig. 1a. The interior angles of the shield are 82° and 158°. On the other hand, Yb atoms connected by the 4Å edges in the $c$-projection of the tetragonal $Yb_2Pd_2Sn$ structure reveal tiling of *squares* and *triangles*. While the triangle tile is just like in the α-YbPdSn centered by a Pd atom at $z$=1/2, the squares are centered by a Sn atom ($z$=1/2) at its center (site $Sn_2$ in Fig. 1b). In this case, edge length of the square is 3.97 Å, and the triangle is not exactly equilateral with two edge lengths of 3.70 Å and 3.97 Å. They correspond respectively to slightly distorted unit cell of *CsCl*-type structure and a half unit cell of $AlB_2$-type structure (see Fig. 89 in ref. 16).

The actual shapes of these three kinds of tiles depend on alloy systems. However we could naturally define ideal tile shapes by their generic symmetries (including their "decoration" - atomic motif bound to them) when they are isolated from the surrounding context: the triangle tile is equilateral with *3m* symmetry, square with *4mm*, while the shield tile has only 3-fold rotation. We further notice that the shield tile can be interpreted as an intergrowth of three square tiles overlapping near their missing corner: it then follows that the ideal shield tile will alternate internal angles of π/2 and 5π/6. Since edge-to-edge assembly of any of the three idealized tiles is entirely plausible (at least in terms of interatomic distances), we hypothesize existence of a phase with *c*=3.72 Å, whose structure can be described in *c*-projection as a combined tiling of all three kinds of tiles, with *a*=4.00 Å edge length, and featuring global 12-fold symmetry (with 3 families of squares, 4 families of triangles and shields, each related by π/6 rotations).

In the following we demonstrate, firstly that a model of such 12-fold symmetric quasicrystal is actually constructed using three kinds of tiles, and secondly that the two crystal structures formed in Yb-Sn-Pd system result from the application of linear phason strain to the quasicrystal model; in other words, the two structures are approximants of an as yet undiscovered dodecagonal quasicrystal.

### 3. Projection method to relate approximants to dodecagonal quasicrystal

Before constructing the structure model of the quasicrystal, the projection method is briefly reviewed. The 3-dimensional dodecagonal quasicrystal is regarded as a periodic stacking of 2-dimensional quasiperiodic tilings along the 12-fold symmetry axis. A structure model of the quasicrystal is obtained from the 5-dimensional periodic structure with two lattice parameters, $a_{DD}$ and $c$ [18-20]. In this method, the lattice points in the 5-dimensional space is first projected onto the 2-dimensional phason space for selection. If the projected point is located inside the acceptance region $W$, or window, defined in the phason space, the corresponding quasilattice point, $r_{//}$, appears in the 3-dimensional physical space.

$$\{r_{//} | r_\perp \in W\}, \quad \text{where} \quad r_{//} = \sum_{n=1}^{5} m_n a_{n//}, \text{ and } \quad r_\perp = \sum_{n=1}^{5} m_n a_{n\perp} \text{ ------ (1).}$$

Here $a_{n//}$ and $a_{n\perp}$ are the physical and the phason components of vector $a_n$, respectively (see Appendix 1 for details). The set of integers $\{m_n\}$ is an index of the lattice point. If one adopts a dodecagon as an acceptance region, presented in Fig. 2a, with a radius of $a_{DD}/\sqrt{2}$, *Shield tiling* appears in the *c*-projection in the physical space. This tiling consists of three kinds of tiles with the common edge length $a=a_{DD}/\sqrt{2}$; the square, the equilateral triangle and the shield [18]. A periodic lattice corresponding to an approximant is obtained from the same 5-dimensional lattice, when the linear phason strain $D$ is applied appropriately [5, 19-21].

$$\{r_{//} | r_{\perp} + Dr_{//} \in W\} \quad \text{------ (2).}$$

The atomic model of a quasicrystal or an approximant can be derived by adding atom positions in the 5-dimensional unit cell, and also by defining the corresponding acceptance regions.

In order to apply the projection method, atomic positions in the two structures, α-YbPdSn and $Yb_2Pd_2Sn$, are indexed using five vectors $a_{n//}$ presented in Fig. 1. For example, all Yb atoms are located at vertices of tiles. They are indexed as a set of integers $\{m_n\}$, and then the representative position is [00000]. The $Pd_1$ atom is located at the center of the equilateral triangle, and the index is the type of [1/3 0 1/3 0 1/2]. Similarly, the $Pd_2$ and $Sn_2$ atoms are indexed as [1/3 1/3 1/3 1/3 0] and [1/2 00 1/2 1/2], respectively. The atomic position of the $Sn_1$ is indexed as [1/2 00 1/2 1/2] by assuming the idealized shape of the shield tile described above. The atomic positions in the 5-dimensional crystal were derived from such indexing. They are summarized in the second column in Table 1. These atomic positions are special points of the 5-dimensional dodecagonal lattice, and correspond to those of 2-dimensional quasilattice classified by Niizeki [22]. The multiplicity of each site is listed in the third column in Table I.

The shape of each acceptance region is also determined referring to the structures of the two approximants. For example, the $Pd_1$ atoms need to appear at all centers of the triangles, and not to appear at other places. Such condition enforces the shape and the size of each acceptance region. The acceptance regions thus obtained are presented in Fig. 2. For example in Fig. 2d, the region for Sn at [1/2 00 1/2 1/2] satisfies 4-fold symmetry, and consists of two parts; the inner square for $Sn_2$ and the surrounding four triangles for $Sn_1$. As a whole, the atomic positions and the corresponding acceptance regions listed in Table I satisfy the point symmetry 12/*mmm*. Then the space group of the 5-dimensional crystal is *P*12/*mmm* [19, 20].

Using these acceptance regions, the 3-dimensional model of a hypothetical dodecagonal quasicrystal is constructed by the projection method (see Fig. 3). It is noted that there is no atomic cluster with local 12-fold symmetry in this model, while many dodecagons with edge length $a$ are embedded. The alloy composition is Yb:Pd:Sn = $\frac{1+\sqrt{3}}{2} : \frac{1+\sqrt{3}}{2} : 1$, which is calculated from the areas of the acceptance regions presented in Table I. This composition is in between α-YbPdSn and $Yb_2Pd_2Sn$.

The structure factors $|F(g_{//})|$ were calculated from the 5-dimensional structure using the standard method [19, 20] and are presented in Fig. 4. Here the atomic scattering factors of Pd, Sn and Yb were taken into account. There is no extinction rule in accordance with the space group *P12/mmm*. The apparent disappearance of 00001 reflection in Fig. 4b is accidental (for reflection index $\{h_n\}$ see Appendix 1). Except for $0000h_5$ type with multiplicity 2, the strongest reflection is 11101 indicated by A in Fig. 4b. This reflection with the lattice spacing $d$=2.58 Å has multiplicity 24, and forms a very strong isolated peak in a powder X-ray diffraction pattern (not shown here).

The ideal structure models of the α-YbPdSn and the $Yb_2Pd_2Sn$ structures can be constructed by introducing the linear phason strains $D_h$ and $D_t$ [21, 23], respectively, in the formula (2);

$$D_h = \frac{1}{2+\sqrt{3}}\begin{pmatrix} 0 & 0 & 0 & 0 & 0 \\ 0 & 0 & 0 & 0 & 0 \\ 0 & -1 & 0 & 0 & 0 \\ 1 & 0 & 0 & 0 & 0 \\ 0 & 0 & 0 & 0 & 0 \end{pmatrix} \text{----- (3),}$$

$$D_t = \frac{1}{4+2\sqrt{3}}\begin{pmatrix} 0 & 0 & 0 & 0 & 0 \\ 0 & 0 & 0 & 0 & 0 \\ 1 & \sqrt{3} & 0 & 0 & 0 \\ \sqrt{3} & -1 & 0 & 0 & 0 \\ 0 & 0 & 0 & 0 & 0 \end{pmatrix} \text{----- (4).}$$

Their derivations are described in Appendix 2. The fact that all atomic positions in the two structures are generated from the common 5-dimensional crystal indicates that they are certainly approximants of the dodecagonal quasicrystal.

## 4. Discussion
### 4.1. Geometrical characteristics of the dodecagonal quasicrystal

In this section geometrical characteristics of the present quasicrystal model will be examined concerning with local atomic arrangement as well as tiling. Comparison of the present model with other dodecagonal quasicrystals may reveal its uniqueness.

In general, the ideal structure of a quasicrystal is understood as a quasiperiodic arrangement of local structural motifs, tiles in this case, with fixed atomic decoration. The *shield tiling* applied in the present model is one of the typical dodecagonal tilings. However, actual dodecagonal quasicrystals known so far (in alloys) belong to different type of tiling; triangle-square tiling almost without other type of tiles. The quasicrystal formed in Mn-Cr-Ni-Si alloy [24] is composed of Frank-Kasper coordination polyhedra; CN12, CN14 and CN 15. In the *c*-projection pseudo 12-fold clusters, corresponding to CN14 polyhedra, are located at every vertices of triangle-square tiling with edge length 4.55 Å. The clusters further form larger dodecagon as shown in Fig. 5b with the diameter 12.4 Å . The chalcogenide $Ta_{1.6}Te$ is another example of triangle-square tiling [25], while this quasicrystal consists of wavy layers stacking along the 12-fold axis [26]. If one views arrangement of Ta atoms along the axis, the projection of the structure is quite similar to that of Mn-based quasicrystal. The edge length is approximately 5.2 Å. Also in this quasicrystal, the larger dodecagonal arrangement formed by Ta and Te atoms can be found, which is again similar to Fig. 5b with the diameter 19.5 Å. Comparing with these known quasicrystals, the present model is remarkably simple without an apparent cluster showing 12-fold symmetry at each vertex of tiling. There are only two layers per repeat with fewest known number of atoms per basic tiles (2 per square, 1.5 per triangle, 6 per shield). This may be a novel class of dodecagonal quasicrystal.

In order to make tiling characteristics clear, let's think of general tiling consisting of square, triangle and shield tiles. Such tiling of the plane will be in one-to-one correspondence with plausible (at least geometrically) Yb-Pd-Sn atomic structure, and is considerably less restrictive than the quasicrystal model we propose. For example, it does not prohibit two squares sharing common side, without the intervening triangle seen always in $Yb_2Pd_2Sn$ tiling. Similarly it does not prohibit shield-shield or square-shield tile pairing. Remarkably, the quasicrystal model generated by the projection method described in Sec. 3 follows very closely tile-tile packing rules seen in the two Yb-Pd-Sn crystals: every square has triangles attached to all four of its edges, forming a *square-star* composed supertile, and every shield shares all of its boundary edges with six triangles, composing *shield-star* supertile. These supertiles are outlined blue in the right bottom part of Fig. 3, and the square-star supertiles are highlighted in light-blue color in Fig. 3. The only spaces left are shield tiles that have six triangles on their edges forming shield-star. In Appendix 3 we prove, that entire quasicrystal structure is *covered* by these two supertiles.

While our dodecagonal tiling satisfies the covering requirement for the two *star* supertiles found in crystal approximants, such covering does not enforce quasiperiodic ordering; indeed it is easy to construct periodic/crystalline coverings by the two supertiles. Intriguingly, entire tiling patch from Fig. 3 is simultaneously covered by a *single* supertile – a dodecagon with a pair of shields inside of it, presented in Fig. 5a. A closely related covering by a single type of dodecagonal patch has been recently proposed by Liao with coworkers [27]; our shield tile always resolves into a square, a pair of triangles and a thin rhombus tile in their model. The possibility of single-cluster covering using the dodecagon patch in Fig. 5a is of high relevance, since energetic favorability of such motif may play role of quasiperiodic order selecting factor [28-30].

In the present structure model based on the shield tiling, the number ratios of three tiles are fixed. However, geometrically, it is possible to reduce shield tiles by combining pair of them into three squares and four triangles: the dodecagonal supertile shown in Fig. 5a is converted to another dodecagon (Fig. 5b) with six squares plus twelve triangles. By this replacement one vertex is created where Yb atom is located. In other words, the shield tile is regarded as a pseudo-vacancy defect. In a general dodecagonal tiling, the number ratio of shield tiles (hence alloy composition, too) is controlled by an extra free parameter.

### 4.2. Metallurgical approach to new type of dodecagonal quasicrystal

The above discussion about the new type of dodecagonal quasicrystal is purely based on geometrical consideration. In this section we will discuss possible approach how to realize it in actual alloys. The primary question is which alloy system we should select for the exploration. In the recent database [17], the hexagonal ZrNiAl-type and tetragonal $Mo_2FeB_2$-type structures can be found in more than 560 and 280 ternary alloy systems, respectively. Among them, both types are formed in more than 120 alloy systems. They are mainly rare-earth alloys containing noble or transition metals as the second component, and main group elements such as Mg, Cd, In, Sn and Pb as the third component. They are listed in Table II. In these alloy systems, two types of structures tend to form as adjacent phases, and these should be candidates for exploration. Further criteria for packing the three kinds of tiles are as follows: periodic repeat along the *c*-axis must be equal or at least approximately equal, namely $c_h \approx c_t$. The edge lengths of tiles are also expected to be equal, and this condition may be examined indirectly by checking equality of lattice parameters $a_h$ and $a_t$, because they are exactly the same in the ideal case. In addition to these conditions, the interior angles of tiles may be important. In Table II the ranges of two ratios, $a_h/a_t$ and $c_h/c_t$, are listed for each alloy system. The ratio of $c_h/c_t$ is always lager than unity, and near-equality can be found in few systems. With respect to the ratio $a_h/a_t$, in some alloys the equality condition is fulfilled, for example in R-(Ni, Cu, Rh, Pd)-In and R-Pd-(Sn, Pb) systems. According to the available information of the hexagonal structures, the obtuse interior angles of the shield tiles range from 149º to 158º that should be 150º in the ideal model. In the case of the tetragonal structures, the interior angles of the isosceles triangles range from 54º to 59º, and they are slightly smaller than the ideal one 60º. One of the candidates satisfying almost the length conditions is Er-Ni-In system with $a_h/a_t = 1.011$ and $c_h/c_t = 1.019$, while the angles are 156.2º and 55.9º. Some local relaxation seems to be necessary to form the dodecagonal quasicrystal.

These matching criteria of tiles are merely local conditions, and for the formation of a quasicrystal the tiles need to be arranged in quasiperiodic manner. Such quasiperiodic structure may be formed as a low-temperature phase, because the long-range order is expected only at low-temperature. The other possibility is formation as an entropy-stabilized phase. Randomness due to phason flip or the replacement shown in Fig. 5 may cause configurational entropy of tiles. In such a case, the quasicrystal is formed possibly as a high-temperature phase or even as a meta-stable phase. Future study exploring this new type of quasicrystal may shed light on long-time question why a quasicrystal is formed.

### 4.3. Relationship between quantum criticality and quasicrystal-related structures

Hereafter we will discuss structural properties of some Yb-based crystals, which suggest hidden relationship between emergence of quantum criticality and formation of structures related to a quasicrystal. This argument is based on curious coincidence in valence-fluctuating crystal/quasicrystal. The keyword is "quasicrystal-related". Examples treated here are α-YbPdSn and $Yb_2Pd_2Sn$, polymorphous α- and β-$YbAlB_4$ crystals, and the icosahedral quasicrystal and its cubic approximant crystal in Au-Al-Yb alloy. As already mentioned in Sec. 1, the former two crystals or the isostructural YbAgGe exhibit intriguing physical properties related to quantum criticality, and they are approximants of a dodecagonal quasicrystal as mentioned above. The Au-Al-Yb approximant and the icosahedral quasicrystal may be the most typical examples of this relationship [6, 7]. With respect to α- and β-$YbAlB_4$, is there any structural relationship to a quasicrystal ?

The possibility of a 7-fold heptagonal quasicrystal has been discussed in the case of Boron-based alloys, and the following crystals were listed as an approximant; $YCrB_4$ (space group: *Pbam*) and $ThMoB_4$ (space group: *Cmmm*) [20, 31]. In their *c*-projection, both structures are regarded as tilings of a unique distorted hexagon with almost equal edge lengths and two interior angles approximately $4\pi/7$ and $6\pi/7$, while the hexagons are arranged in different manner in the two structures. The hexagon can be related to two-dimensional heptagonal tiling, Fig. 1 in [32], consisting of three kinds of rhombuses with equal edge lengths and acute angles $\pi/7$, $2\pi/7$ and $3\pi/7$ (see also Fig. 1 in [31]). The heavy fermion α- and β-YbAlB$_4$ are isostructures to YCrB$_4$ and ThMoB$_4$, respectively. In β-YbAlB$_4$ structure [33, 34], Yb atoms at $z=0$ form the distorted hexagon, and seven-member ring consisting of B atoms ($z=1/2$) is centered at each vertex of the hexagon in the $c$-projection. The interatomic distances of B atoms range between 1.73 and 1.89 Å, and the seven-member ring is regarded as a distorted heptagon. The structure of α-YbAlB$_4$ has similar feature. Then, α- and β-YbAlB$_4$ are regarded as approximants of a heptagonal quasicrystal. It is noted that B atoms also form pentagonal arrangements, and the interpretation as approximants of a decagonal quasicrystal is also possible [35]. The β-YbAlB$_4$ is remarkable crystal exhibiting quantum criticality at ambient pressure without *tuning*, namely no application of pressure or magnetic field, or no addition of minor element [33]. This feature is very unique and is observed also in Au-Al-Yb quasicrystal, but in no other crystal.

For the present moment, the origin of this curious coincidence is not known. There may be several reasons depending on individual systems; special local environments of Yb atoms (related to pseudo symmetric cluster for example), their variety in complex structures, or proximity to long-range quasiperiodic order. The former two may be important with respect to hybridization between 4$f$- and conduction electrons as discussed in [36]. In order to investigate such relationship between structures and physical properties, the present structure model of a dodecagonal quasicrystal may have advantage. Three kinds of tiles have very simple decoration of atoms, and the gradual approach from simple structure (lower order approximant) to the perfect quasicrystal is possible by applying linear phason mechanism. By this approach, the role of critical character of wave functions in a quasperiodic system may be clarified.

## 5. Conclusion

We have demonstrated that α-YbPdSn and Yb$_2$Pd$_2$Sn structures can be regarded as approximants of the hypothetical dodecagonal quasicrystal. The Yb-Pd-Sn alloy system treated here is merely an example, and the same argument is shared among as many as 120 ternary alloy systems with stable ZrNiAl- and Mo$_2$FeB$_2$-type phases [17]. The quasicrystal combining structural motifs from these crystal structures in a subtle quasiperiodic manner may provide new clues in search for genuine causes of the relation between emergence of the quantum criticality and the formation of the quasicrystal-related structures.

**Acknowledgements**

The authors thank Kotoba Toyonaga and Yuuki Hirashima for their contributions to this research project. This work was supported by JSPS KAKENHI Grant Number 17K05524. MM acknowledges support from VEGA 2/0082/17.

**References**

[1] V. Elser and C.L. Henley, Phys. Rev. Lett., **55** (1985) p.2883.

[2] M. Duneau and A. Katz, Phys. Rev. Lett., **54** (1985) p.2688.

[3] P.A. Kalugin, A.Y. Kitaev and L.S. Levitov, JETP Lett., **41** (1985) p.145.

[4] V, Elser, Acta Crystallogr. A, **42** (1986) p.36.

[5] J.E.S Socolar, T.C. Lubensky and P.J. Steinhardt, Phys. Rev., **B34** (1986) p.3345.

[6] K. Deguchi, N.K. Sato, T. Hattori, K. Ishida, H. Takakura and T. Ishimasa, Nature Materials, **11** (2012) p.1013.

[7] S. Matsukawa, K. Deguchi, K. Imura, T. Ishimasa and N.K. Sato, J. Phys. Soc. Jpn., **85** (2016) 063706.

[8] M. Giovannini, R. Pasero and A. Saccone, Intermetallics, **18** (2010) p.429.

[9] F. Gastaldo, M. Giovannini, A.M. Strydom, R.F. Djoumessi, I. Čurlík, M. Reiffers, P. Solokha and A. Saccone, J. Alloys Comp., **694** (2017) p.185.

[10] D. Kußmann, R. Pöttgen, B. Künnen and G. Kotzyba, Z. Kristallogr., **213** (1998) p.356.

[11] T. Görlach, S. Putselyk, A. Hamann, T. Tomanić, M. Uhlarz, F.M. Schappacher, R. Pöttgen and H.v. Löhneysen, Phys. Rev., **B76** (2007) 205112.

[12] Y. Tokiwa, M. Garst, P. Gegenwart, S.L. Bud'ko and P.C. Canfield, Phys. Rev. Lett., **111** (2013) 116401.

[13] T. Muramatsu, T. Kanemasa, T. Kagayama, K. Shimizu, Y. Aoki, H. Sato, M. Giovannini, P. Bonville, V. Zlatic, I. Aviani, R. Khasanov, C. Rusu, A. Amato, K. Mydeen, M. Nicklas, H. Michor and E. Bauer, Phys. Rev., **B83** (2011) 180404(R).

[14] E. Bauer, G. Hilscher, H. Michor, Ch. Paul, Y. Aoki, H. Sato, D.T. Adroja, J.-G. Park, P. Bonville, C. Godart, J. Sereni, M. Giovannini and A. Saccone, J. Phys.: Condens. Matter, **17** (2005) S999.

[15] M.S. Kim and M.C. Aronson, Phys. Rev. Lett., **110** (2013) 017201.

[16] Y.M. Kalychak, V.I. Zaremba, R. Pöttgen, M. Lukachuk and R.-D. Hoffmann, *Handbook on the Physics and Chemistry of Rare Earths*, Vol. 34, edited by K.A. Gschneidner Jr., J.-C.G. Bünzli and V.K. Pecharsky, Elsevier B.V., Amsterdam, 2005, Chap. 218.

[17] P. Villars, K. Cenzual and R. Gladyshevskii, *Handbook of Inorganic Substances 2017*, Walter de Gruyter Belin/Boston, 2017.

[18] F. Gähler, Doctor thesis of ETH-Zürich, Diss. ETH No. 8414 (1988), http://e-collection.library.ethz.ch/view/eth:37444.

[19] A. Yamamoto, Acta Crystallogr., A, **52** (1996) p. 509.

[20] W. Steurer and S. Deloud, *Crystallography of Quasicrystals*, Springer Series in Materials Science 126, Springer-Verlag Belrin Heidelberg, 2009.

[21] P.W. Leung, C.L. Henley and G.V. Chester, Phys. Rev., **B39** (1989) p. 446.

[22] K. Niizeki, J. Phys. A: Math. Gen., **22** (1989) p. 4281.

[23] C. Xiao, N. Fujita, K. Miyasaka, Y. Sakamoto and O. Terasaki, Nature, **487** (2012) p. 349.

[24] T. Ishimasa, S. Iwami, N. Sakaguchi, R. Oota and M. Mihalkovič, Phil. Mag., **95** (2015) p. 3745.

[25] M. Conrad, F. Krumeich and B. Harbrecht, Angew. Chem. Int. Ed., **37** (1998) p. 1384.

[26] M. Conrad and B. Harbrecht, Chem. Eur. J., **8** (2002) p. 3094.

[27] L.G. Liao, W.B. Zhang, T.X. Yu and Z.X. Cao, Chin. Phys. Lett., **30** (2013) 026102.

[28] P. Gummelt, Geometriae Dedicate, **62** (1996) p. 1.

[29] P.J. Steinhardt, H.-C. Jeong, K. Saitoh, M. Tanaka, E. Abe and A.P. Tsai, Nature, **396** (1998) p. 55.

[30] F. Gähler, Mater. Sci. Eng. A, **294-296** (2000) p. 199.

[31] W. Steurer, Phil. Mag., **87** (2007) p. 2707.

[32] P. Repetowicz and J. Wolny, J. Phys. A: Math. Gen., **31** (1998) p. 6873.

[33] Y. Matsumoto, S. Nakatsuji, K. Kuga, Y. Karaki, N. Horie, Y. Shimura, T. Sakakibara, A.H. Nevidomskyy and P. Coleman, Science, **331** (2011) p. 316.

[34] K. Kuga, G. Morrison, L. Treadwell, J.Y. Chan and S. Nakatsuji, Phys. Rev. B, **86** (2012) 224413.

[35] M. Mihalkovič and M. Widom, Phys. Rev. Lett., **93** (2004) 095507.

[36] S. Watanabe and K. Miyake, J. Phys. Soc. Jpn., **82** (2013) 083704.

**Appendix 1**

Several schemes have been used for indexing of a dodecagonal quasicrystal [18-24]. In these schemes, the parameter sets, $\alpha = \pm\frac{5\pi}{6}$ and $\beta = \pm 1$ (any double sign), are used in the expression of the dodecagonal lattice in the 5-dimensional space*, where the orthonormal basis is used. In this paper, the parameter set, $\alpha = -\frac{5\pi}{6}$ and $\beta = -1$, is used.

$$\boldsymbol{a}_n = \frac{a_{DD}}{\sqrt{2}} \begin{pmatrix} \cos\frac{\pi}{6} & -\sin\frac{\pi}{6} & 0 & 0 & 0 \\ \sin\frac{\pi}{6} & \cos\frac{\pi}{6} & 0 & 0 & 0 \\ 0 & 0 & \cos\alpha & -\sin\alpha & 0 \\ 0 & 0 & \sin\alpha & \cos\alpha & 0 \\ 0 & 0 & 0 & 0 & 1 \end{pmatrix}^{n-1} \begin{pmatrix} 1 \\ 0 \\ \beta \\ 0 \\ 0 \end{pmatrix} \quad \text{for } n=1\sim 4,$$

$$\boldsymbol{a}_5 = c \begin{pmatrix} 0 \\ 0 \\ 0 \\ 0 \\ 1 \end{pmatrix}.$$

Here $a_{DD}$ and $c$ denote the lattice parameters of the 5-dimensional lattice. The parameter $a_{DD}$ is related to the common edge length, $a$, of the square, equilateral triangle and shield as follows;

$$a = \frac{a_{DD}}{\sqrt{2}}.$$

The parameter $c$ denotes periodicity along the 12-fold axis. The physical components of the lattice vectors are expressed as follows;

$$\boldsymbol{a}_{1//} = a\boldsymbol{e}_1, \qquad \boldsymbol{a}_{2//} = \frac{a}{2}\left(\sqrt{3}\boldsymbol{e}_1 + \boldsymbol{e}_2\right),$$

$$\boldsymbol{a}_{3//} = \frac{a}{2}\left(\boldsymbol{e}_1 + \sqrt{3}\boldsymbol{e}_2\right), \quad \boldsymbol{a}_{4//} = a\boldsymbol{e}_2, \quad \boldsymbol{a}_{5//} = c\boldsymbol{e}_5.$$

Their phason components are

$$\boldsymbol{a}_{1\perp} = -a\boldsymbol{e}_3, \qquad \boldsymbol{a}_{2\perp} = \frac{a}{2}\left(\sqrt{3}\boldsymbol{e}_3 + \boldsymbol{e}_4\right),$$

$$\boldsymbol{a}_{3\perp} = \frac{-a}{2}\left(\boldsymbol{e}_3 + \sqrt{3}\boldsymbol{e}_4\right), \quad \boldsymbol{a}_{4\perp} = a\boldsymbol{e}_4, \quad \boldsymbol{a}_{5\perp} = 0.$$

Here $\boldsymbol{e}_n$ ($n = 1\sim 5$) denote orthonormal basis. The three vectors $\boldsymbol{e}_1$, $\boldsymbol{e}_2$ and $\boldsymbol{e}_5$ are embedded in the 3-dimensional physical space and the rest two vectors in the 2-dimensional phason space. These vectors are illustrated in Figs. 1 and 2. Then, the lattice vectors $\boldsymbol{a}_n$ of the 5-dimensional dodecagonal lattice are expressed as follows by using the matrix $M$.

---

*: In some cases, similar expression is used for the definition of the reciprocal lattice vectors $\boldsymbol{a}_n^*$ in stead of the lattice vectors $\boldsymbol{a}_n$.

$$\{a_n\} = M\{e_n\}, \quad \text{where} \quad M = \frac{a}{2}\begin{pmatrix} 2 & 0 & -2 & 0 & 0 \\ \sqrt{3} & 1 & \sqrt{3} & 1 & 0 \\ 1 & \sqrt{3} & -1 & -\sqrt{3} & 0 \\ 0 & 2 & 0 & 2 & 0 \\ 0 & 0 & 0 & 0 & 2c/a \end{pmatrix}.$$

The reciprocal vectors are obtained by calculating the inverse matrix of $M^T$, where $M^T$ denotes the transverse matrix of $M$.

$$\{a_n^*\} = M^{T^{-1}}\{e_n\}.$$

Thus, the physical components of the reciprocal lattice vectors illustrated in Fig. 4 are

$$a_{1//}^* = \frac{1}{2\sqrt{3}a}(\sqrt{3}e_1 - e_2), \qquad a_{2//}^* = \frac{1}{\sqrt{3}a}e_1,$$

$$a_{3//}^* = \frac{1}{\sqrt{3}a}e_2, \quad a_{4//}^* = \frac{1}{2\sqrt{3}a}(-e_1 + \sqrt{3}e_2).$$

$$a_{5//}^* = \frac{1}{c}e_5.$$

The reflection of the quasicrystal is indexed using five integers $h_n$, $n=1\sim5$, as follows;

$$g_{//} = \sum_{n=1}^{5} h_n a_{n//}^*.$$

**Appendix 2**

The linear phason strains, $D_h$ and $D_t$ in (3) and (4), are derived as follows. In the case of hexagonal α-YbPdSn, the lattice vectors $a_h$ and $b_h$ are expressed;

$$a_h = a_{1//} + a_{2//} - a_{4//} = \frac{a}{2}\{(2+\sqrt{3})e_1 - e_2\},$$

$$b_h = -a_{1//} + a_{3//} + a_{4//} = \frac{a}{2}\{-e_1 + (2+\sqrt{3})e_2\}.$$

Corresponding vectors, $a_{h\perp}$ and $b_{h\perp}$, in the phason space are

$$a_{h\perp} = a_{1\perp} + a_{2\perp} - a_{4\perp} = \frac{a}{2}\{(-2+\sqrt{3})e_3 - e_4\},$$

$$b_{h\perp} = -a_{1\perp} + a_{3\perp} + a_{4\perp} = \frac{a}{2}\{e_3 + (2-\sqrt{3})e_4\}.$$

In order to generate the periodic structure, the linear phason strain $D_h$ needs to compensate the displacement in the phason space, and the matrix $D_h$ should satisfy the following conditions;

$$a_{h\perp} + D_h a_{h//} = 0 \quad \text{and} \quad b_{h\perp} + D_h b_{h//} = 0.$$

The matrix $D_h$ in (3) is derived from these conditions. Similarly the phason matrix $D_t$ was obtained as (4) for the tetragonal $Yb_2Pd_2Sn$. In general, lattice vectors ***a*** and ***b*** of an approximant can be expressed by two sets of integral coefficients of the four vectors $\boldsymbol{a}_{n//}$, $n=1\sim4$. Such sets are useful to denote each approximant; $[110\bar{1}]$ and $[\bar{1}011]$ for the hexagonal α-YbPdSn for example.

**Appendix 3**

The Shield tiling presented in Fig. 3 is regarded as covering with two kinds of supertiles, *square-star* and *shield-star*. This covering interpretation is proved from the following properties of the shield tiling by applying one-to-one correspondence between the physical and the phason spaces;

1. A square in the physical space appears invariably as a part of a square-star. In other words, it is always accompanied by four surrounding triangles. This can be proved from the fact that the acceptance region for the square-star is just a square in the phason space exactly the same as that for the simple square (see the central square in Fig. 2d).

2. An equilateral triangle in the physical space has at least one square attached on its three edges. The acceptance region for the triangle is a shield in the phason space (see Fig. 2b). This acceptance region is divided into three parts as presented in Fig. 6. The central region painted orange corresponds to equilateral triangles with three attached squares. The outer (blue) and middle (green) regions correspond to those with single and two squares, respectively. Considering the above property 1, all equilateral triangles must have at least one parent square-star.

From these properties, it is shown that if one paints all square-stars in the shield tiling as in Fig. 3, only one type of tiles, namely the shield, remains unpainted. Furthermore, it is proved that a shield in the physical space appears invariably as a part of a shield-star. Accordingly, the shield tiling is *covering* with two supertiles, the square-star and the shield-star.

**Figure Captions**

Figure 1

Structure models of approximants. a: hexagonal α-YbPdSn, b: tetragonal $Yb_2Pd_2Sn$. Red, yellow and green colored circles denote Yb, Pd and Sn atoms, respectively. Four vectors $a_{1//}$ ~ $a_{4//}$ are shown in the figure. Vector $a_{5//}$ is perpendicular to the plane of this figure.

Figure 2

Four acceptance regions in the 5-dimensional structure model of the dodecagonal quasicrystal. a: acceptance region for Yb atoms located at [00000], b: $Pd_1$ atoms located at [1/3 0 1/3 0 1/2], c: $Pd_2$ atoms located at [1/3 1/3 1/3 1/3 0] and d: $Sn_1$ and $Sn_2$ atoms located at [1/2 00 1/2 1/2]. Centered square and surrounding four triangles correspond to $Sn_2$ and $Sn_1$, respectively.

Figure 3

Structure model of hypothetical dodecagonal quasicrystal viewed along the 12-fold axis. Two types of supertiles, square-star and shield-star, are outlined in the lower part of the figure. All possible square-stars are highlighted in light-blue color. Common edge length is $a$=4.00 Å. For the $z$-component of each atom, see text.

Figure 4

Calculated structure factors of the dodecagonal quasicrystal. a: the plane at $h_5$=0. b: $h_5$=1. Strong reflection indicated by A in b has index 11101.

Figure 5

Two types of dodecagons with different internal tilings. The number of vertices included in a dodecagon is increased from six in (a) to seven in (b) by the tiling conversion.

Figure 6 (in Appendix 3)

Subdivision of the acceptance region for equilateral triangle. The region with shield shape is divided into three parts according to the number of squares surrounding a triangle.

Table I

Acceptance regions of the dodecagonal quasicrystal model. Areas in the fifth and sixth columns are normalized by the factor of $a^2$. *: a set of four triangles surrounding the square centered at [1/2 00 1/2 1/2].

| Site | Position | Multiplicity | Shape | Area | Area · Multiplicity |
|------|----------|--------------|-------|------|---------------------|
| Yb | [00000] | 1 | Dodecagon | 3 | 3 |
| $Pd_1$ | [1/3 0 1/3 0 1/2] | 4 | 3-fold hexagon | $(3-\sqrt{3})/2$ | $6-2\sqrt{3}$ |
| $Pd_2$ | [1/3 1/3 1/3 1/3 0] | 4 | Triangle | $(2\sqrt{3}-3)/4$ | $2\sqrt{3}-3$ |
| $Sn_1$ | [1/2 00 1/2 1/2] | 3 | Four triangles* | $2\sqrt{3}-3$ | $6\sqrt{3}-9$ |
| $Sn_2$ | [1/2 00 1/2 1/2] | 3 | Square | $2-\sqrt{3}$ | $6-3\sqrt{3}$ |

Table II

Rare-earth alloys forming both hexagonal ZrNiAl- and tetragonal $Mo_2FeB_2$-type structures. Ranges of ratios of lattice parameters, $a_h/a_t$ and $c_h/c_t$, in each system are listed in the third and fourth columns, respectively.

| system | R: rare-earth element | $a_h/a_t$ | $c_h/c_t$ |
|---|---|---|---|
| R-Cu-Mg | Y, La, Ce | 0.974-0.976 | 1.057-1.075 |
| R-Pd-Mg | Y, Ce, Pr, Nd, Sm, Gd, Tb, Dy, Ho | 0.970-0.987 | 1.026-1.072 |
| R-Cu-Cd | Sm, Gd, Dy | 0.987-0.989 | 1.047-1.051 |
| R-Pd-Cd | La, Ce, Nd, Gd | 0.987-0.989 | 1.034-1.040 |
| R-Au-Cd | La, Ce, Nd, Pr, Sm | 0.972-0.978 | 1.039-1.044 |
| R-Ni-In | Y, La, Ce, Pr, Nd, Sm, Gd, Tb, Dy, Ho, Er, Tm, Lu | 0.997-1.022 | 1.016-1.068 |
| R-Cu-In | Y, La, Ce, Pr, Nd, Sm, Gd, Tb, Dy, Ho, Er, Tm, Lu | 0.967-1.003 | 1.033-1.079 |
| R-Rh-In | La, Ce, Pr, Nd | 0.975-1.003 | 1.028-1.075 |
| R-Pd-In | Y, La, Ce, Pr, Nd, Sm, Gd, Tb, Dy, Ho, Er, Tm, Yb, Lu | 0.983-1.004 | 1.027-1.081 |
| R-Pt-In | Y, La, Ce, Pr, Nd, Sm, Gd, Tb, Dy, Ho | 0.983-0.994 | 1.043-1.058 |
| R-Au-In | Y, La, Ce, Sm, Gd, Tb, Dy, Ho, Er, Yb | 0.954-0.985 | 1.036-1.082 |
| R-Pd-Sn | Ce, Yb | 0.978-1.000 | 1.040-1.060 |
| R-Pd-Pb | Y, La, Ce, Pr, Nd, Sm, Gd, Tb, Dy, Ho, Er, Tm | 0.979-1.001 | 1.043-1.072 |
| R-Pt-Pb | La, Ce, Pr, Nd, Sm | 0.972-0.975 | 1.079-1.087 |

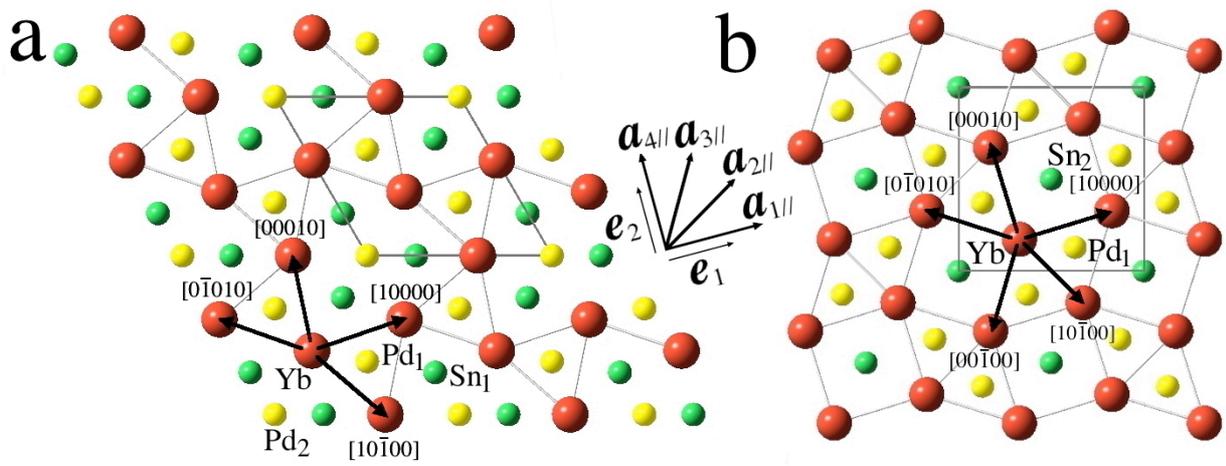

Fig. 1

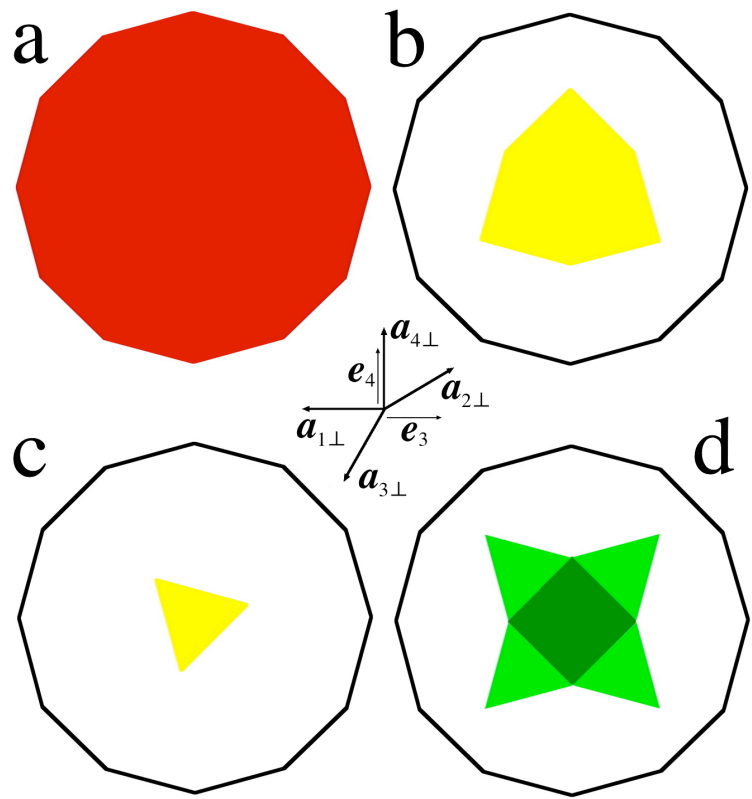

Fig. 2

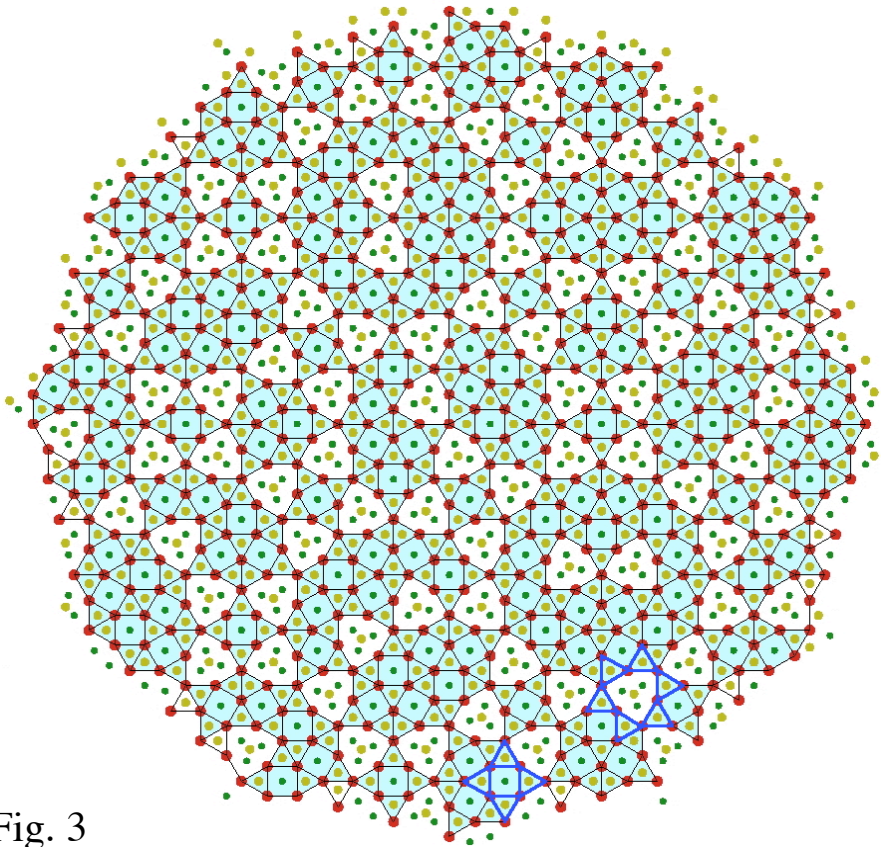

Fig. 3

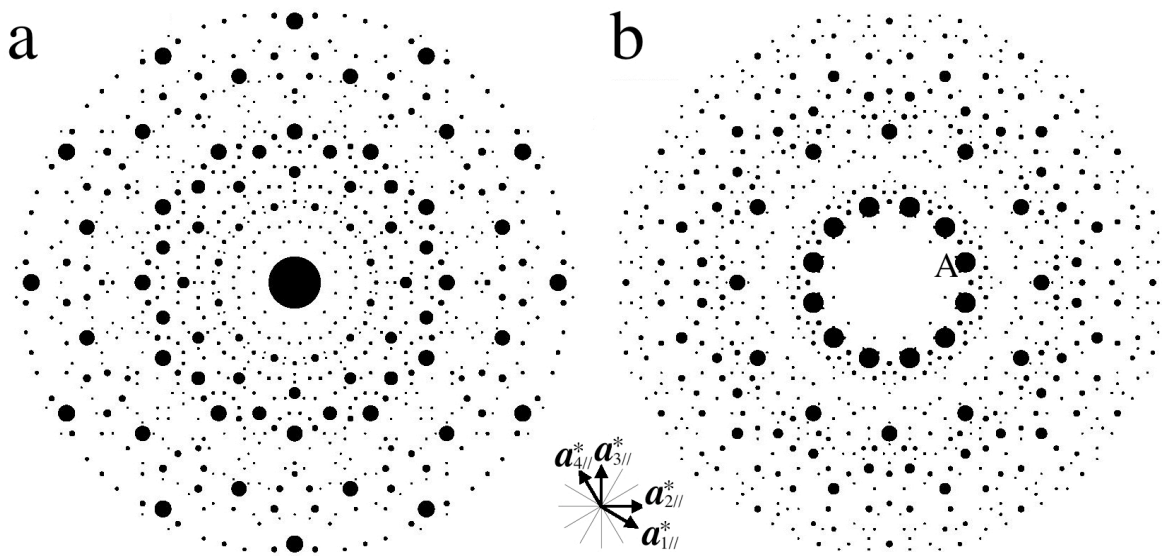

Fig. 4

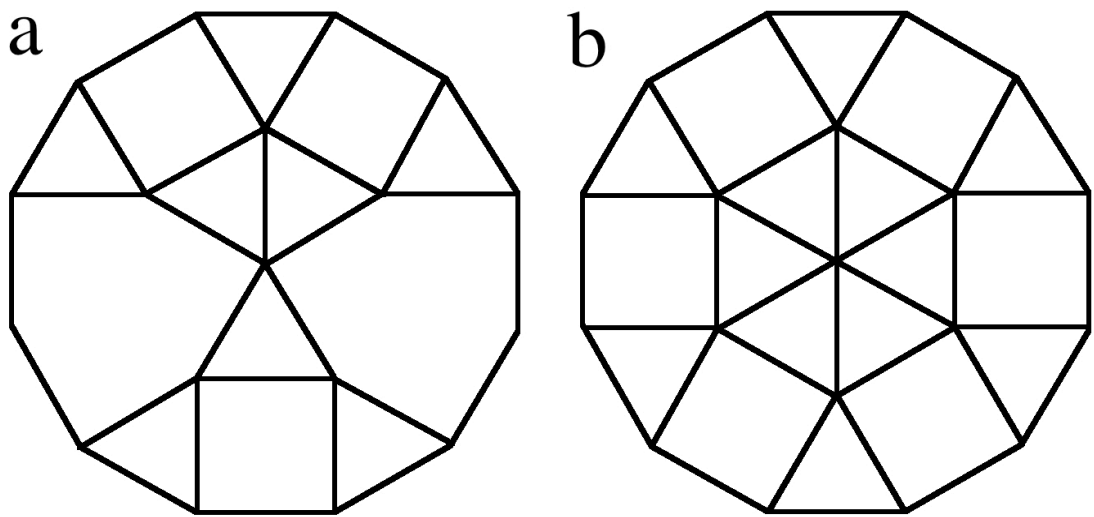

Fig. 5

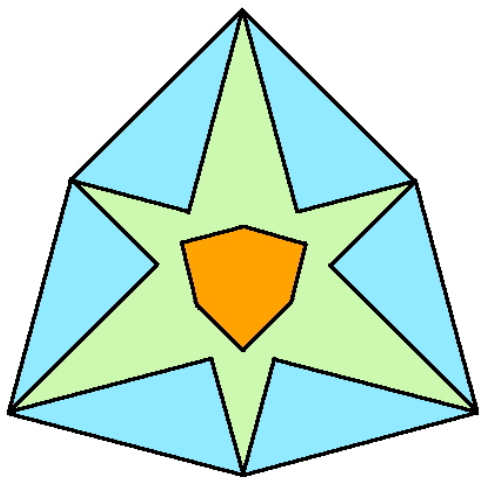

Fig. 6